\documentclass[reqno,11pt]{article}
\pdfoutput=1
\usepackage{jheppub}
\usepackage{epsfig}
\usepackage{amssymb}
\usepackage{verbatim}
\usepackage{amsmath}
\usepackage{relsize}
\usepackage{hyperref}
\usepackage{dsfont}
\usepackage{slashed}
\usepackage[dvipsnames]{xcolor}
\usepackage{epstopdf}
\usepackage{graphicx}
\usepackage{graphicx}
\usepackage{caption}
\usepackage{subcaption}
\usepackage{mathtools}
\def\nn{\nonumber}
\def\bz{\bar{z}}
\def\e{{\rm e}}

\def\b{\beta}

\newcommand{\be}{\begin{equation}}
\newcommand{\ee}{\end{equation}}
\newcommand{\ba}{\begin{align}}
\newcommand{\ea}{\end{align}}
\newcommand{\bea}{\begin{eqnarray}}
\newcommand{\eea}{\end{eqnarray}}
\newcommand{\bc}{\begin{center}}
\newcommand{\ec}{\end{center}}

\title{OPE inversion in Mellin space}
\author[a]{Carlos Cardona}

\affiliation[a]{
Niels Bohr International Academy and Discovery Center,\\University of Copenhagen, Niels Bohr Institute\\ Blegamsvej 17, DK-2100 Copenhagen Ø, Denmark}
\emailAdd{carlosgiraldo@nbi.ku.dk}

\abstract{The fundamental ingredients that build the observables in conformal field theory are the spectrum of operators and the OPE coefficients, or equivalently, the two and three-point functions of the theory. Recently an inversion formula  solving the OPE coefficients by a convolution over the light-cone double-discontinuities of the correlator has been found by Simon Caron-Huot. Taking into account that the same OPE data determine the Mellin amplitude representation of the correlator, motivate us to look for an analogous inversion formula in Mellin space, which we develops partially on this paper.}

\begin{document}

\maketitle 

\section{Introduction}\label{sect:intro}
The basic building blocks of correlation functions of local operators in conformal field theories are given by the spectrum and the OPE coefficients. Despite intense study of interacting conformal field theories over the last couple of decades, we still lack of a general framework that allow us to compute those fundamental blocks from first principles and possibly at any coupling. The most promising candidate so far for such a framework is the conformal bootstrap program, which adheres to the ideal that physical observables, such as correlation function in conformal field theories, should be constrained or even solved by imposing on them a minimal set of physical requirements, such as unitarity, locality, crossing, space-time symmetries and perhaps a general set of inner symmetries.

On the CFT department of this program, there has been a successful resurgence of this idea involving many new intriguing and exciting results. From the numerical approach, tremendous progress has been made mostly boosted by the techniques developed in \cite{Rattazzi:2008pe}, where the problem of bounding operator dimensions by crossing symmetry conditions was revisited. Among the most popular subsequent  applications of this new techniques concerns to the famous 3D Ising model \cite{ElShowk:2012ht, Kos:2014bka, El-Showk:2014dwa, Simmons-Duffin:2016wlq}. The amount of work in this direction is so vast that we will not even try to summarize it in this introduction, but instead we refer the reader to some nice updated reviews on the topic \cite{Rychkov:2016iqz,Simmons-Duffin:2016gjk}. 

On the analytic side of the story, the recent progress is as exciting and impressive. By studying the crossing equation of a four point correlation function of scalars as an expansion in inverse powers of the spin for large spin exchanges \cite{Komargodski:2012ek,Fitzpatrick:2012yx,Kaviraj:2015cxa,Kaviraj:2015xsa,Alday:2015eya, Alday:2015ota}, it has been developed a ``perturbation theory on spin"\cite{Alday:2016njk} that can be applied equally at weak and strong coupling regimes. This has been used successfully on a number of examples, remarkably giving accurate results even for operators of spin as large as (or better, as small as) two \cite{Simmons-Duffin:2016wlq, Alday:2015ota}.  Another similar limit that allows to constraint the OPE coefficients is the Regge limit at high energy scattering where some interesting progress has also been made \cite{Costa:2012cb, Costa:2017twz}. In applications to AdS/CFT duality, more particularly to ${\cal N}=4$ SYM, the analytic boostrap have been used recently to compute the OPE coefficients of operators in the stress-tensor multiplet  \cite{Aharony:2016dwx, Alday:2017xua, Alday:2017vkk} which lately should corresponds to loop corrections in AdS space, where not much is know so far, although some modest progress have been done lately \cite{Penedones:2010ue, Giombi:2017hpr, Yuan:2018qva, Yuan:2017vgp, Cardona:2017tsw}

In principle, boostrap techniques based on crossing symmetry are not exclusive for conformal field theories and it should be possible to extend them to constraint S-matrix elements of general field theories. Some important considerations on that regard were made lately as well \cite{Paulos:2017fhb,Paulos:2016but,Paulos:2016fap}

Even more recently, Simon Caron-Huot  developed the CFT analogous of the Froissart-Gribov formula for partial wave S-matrix expansion,  which inverts the conformal block expansion in cross-ratios space and allows to write the OPE coefficients in terms of a convolution over the double discontinuities of the four point correlation function, with a kernel corresponding to a particular conformal block (in practice,  the collinear limit of it) \cite{Caron-Huot:2017vep}. Explicitly the inversion formula looks like,
\be\label{simonformula}
c^t_{\Delta,J}={\kappa_{\beta}\over 4\pi}\int_0^1dz d\bz\mu(z,\bz)G_{\Delta+1-d}^{(J+d-1)}(z,\bz)\,{\rm dDisc}\left[{\cal G}(z,\bz)\right]\,,
\ee
where ${\rm dDisc}\left[{\cal G}(z,\bz)\right]$ denotes the double discontinuity on Lorentz signature arising by crossing the branch cuts from the light-cone singularities, $\mu$ is a measure factor and the coordinates $(z,\bz)$ are esencially the cross-ratios. This formula was re-derived in coordinate space in \cite{Simmons-Duffin:2017nub} where instead of using the Froissart-Gribov trick, the authors used the orthogonality relation between partial waves and its shadows to invert the euclidean OPE to later Wick rotate to Lorentz signature in order to expose the light-cone branch cuts on which the double discontinuities arise \footnote{See also \cite{Raben:2018sjl}, where is also shown how the Lorentzian blocks can be derived directly by inspection of the boundary conditions of the conformal Casimirs}.

Equation \eqref{simonformula} is the main motivation of this short note. Here we take some steps in formulating an analogous inversion formula in Mellin space. As we will see in the main body of this work, it is possible to invert the conformal block expansion in Mellin space by using ortogonality relations among the Mellin residues. The Mellin amplitude representation of the collinear limit of a four point function can be written in terms of almost-Hanh polynomials \cite{Korchemsky:1994um,Costa:2012cb, Dey:2017fab} which satisfy the desired orthogonality. It is then expected, than the full Mellin representation of a given four-point function preserve this ortogonality property, since it can be written in terms of Mack Polynomials \cite{Mack:2009mi} which likely should form an orthogonal basis\footnote{To my knowledge, up today there is not a prove for the orthogonality of Mack polynomials.}.

Even in using a fomula such as \eqref{simonformula} in cross-ratios space, one can see that the Mellin representation have the advantage that the cross-ratios dependence is through simple monomials and hence it is expected that the computation of discontinuities would be easier to perform in Mellin space than over more  complicated functions of the cross-ratios.  This fact has been exploited before for example in \cite{Belitsky:2013xxa}. 
The double discontinuity across the light-cone branch cuts in \eqref{simonformula} is responsible for killing the double twist contributions to the OPE coefficient. In Mellin space it translates to introducing zeros on the Mellin variables at the right positions to cancel the poles associated to the double twist operators.

Even thought it might be possible that a Froissart-Gribov type of contour deformation which introduces the proper zeros  exist purely in Mellin space, in this work we are not intended to study those deformations, but rather are going to use the doble-discontinuity as an input.


Finally, let us end this introduction by highlight some important applications of Mellin space in conformal field theory. It has been remarkable useful and enlightening the application of it in taking the Regge limit on CFT correlation functions \cite{Costa:2012cb, Costa:2017twz}, as well as in taking advantadge of the S-matrix character of the Mellin amplitudes to extract information about correlation functions \cite{Goncalves:2014rfa, Fitzpatrick:2011ia}, or even in some attempts to do the inverse trick and instead use the correlator knowledge to gain information on the S-matrix \cite{Cardona:2016ymb, Cardona:2017keg}. Also very recently a Mellin representation for half-BPS four-point functions in ${\cal N}=4$ SYM has been found in \cite{Rastelli:2017udc,Rastelli:2016nze}. More aligned to the ideas of this paper, a  very interesting boostrap approach in Mellin space has been developed in the works \cite{Gopakumar:2016wkt, Dey:2017fab,Dey:2016mcs,Gopakumar:2016cpb}\footnote{See also \cite{Golden:2017fip} where the Mellin bootstrap has been applied to study interacting theories in $d>6$.}, where unlike here, crossing symmetry is guaranteed by construction and the bootstrap equations are given by imposing the conditions that kill double twist contributions instead.    

\section{Mellin block expansion}
We will consider a conformal correlation function of four scalar primary operators. Conformal invariance dictates that up to a fix prefactor, it is given only  by a function of the cross ratios,
\be
u=z\bz={x_{12}^2x_{34}^2\over x_{13}^2x_{24}^2},\,~v=(1-z)(1-\bz)={x_{23}^2x_{14}^2\over x_{13}^2x_{24}^2}\,,
\ee
as,
\be\label{4pexchange}
\left\langle\prod_{i=1}^4\phi_i(x_i)\right\rangle={1\over (x_{12}^2)^{\Delta_1+\Delta_2\over 2}(x_{34}^2)^{\Delta_3+\Delta_4\over 2}}\left({x_{24}^2\over x_{14}^2}\right)^a\left({x_{14}^2\over x_{13}^2}\right)^b\,{\cal G}(u,v)\,.
\ee
with $a={\Delta_1-\Delta_2\over2}$ and  $b={\Delta_3-\Delta_4\over2}$. The function ${\cal G}(u,v)$ can be expanded in terms of known conformal blocks \cite{Dolan:2000ut},
\be\label{conformalblocksexp}
{\cal G}(u,v)=\sum_{\Delta,J} c_{\Delta,J}\, G^{(J)}_{\Delta}(u,v)\,.
\ee 
Which in even space-time dimension can be written as an expansion in terms of hypergeometric functions. For $d=4$ we have,
\be\label{conformalblocks}
 G^{(J)}_{\Delta}(u,v)={z\bz\over \bz-z}\left({\over}k_{\Delta-J-2}(z)k_{\Delta+J}(\bz)-k_{\Delta+J}(z)k_{\Delta-J-2}(\bz)\right)\,,
\ee
with 
\be\label{collinearblocks}
k_{\beta}(z)=z^{\beta/2}\, {}_2F_1\left({\beta\over 2}+a,{\beta\over 2}+b;\,\beta\,|z\right)\,.
\ee
We want to utilize a Mellin representation for the conformal blocks \cite{Mack:2009mi}, to do so we use the conventions in \cite{Dolan:2011dv} to write,
\bea\label{mellinrepG}
{\cal G}(u,v)= \int^{\rho-i\infty}_{\rho+i\infty}{ds\,dt\over(2\pi i)^2}&\Gamma(-t)\Gamma(-t-a-b)\Gamma(s+t+a)\Gamma(s+t+b)\nn\\
&\Gamma\left({\tau\over2}-s\right)\Gamma\left({d-\beta\over2}-s\right)~{\cal M}(s,t)\,u^s\,v^{-(s+t)}\,,
\eea
where ${\cal M}(s,t)$ is known as the Mellin amplitude and in the context of conformal field theory was introduced by Mack in  \cite{Mack:2009mi}. By comparing \eqref{mellinrepG} to the conformal block expansion \eqref{conformalblocksexp} we can conclude that it should exist an equivalent expansion in Mellin space as,
\be\label{mellinblockexp}
{\cal M}(s,t) =\sum_{\Delta,J} c_{\Delta,J}\,m^{(J)}(s,t)\,,
\ee
such as the transformation of the sub-amplitudes $m^{(J)}(s,t)$ correspond to the usual conformal blocks in cross-ratios $g^{(J)}_{\Delta}(u,v)$, explicitly,
\bea\label{mellinrepg}
g^{(J)}_{\Delta}(u,v)=\int^{\rho-i\infty}_{\rho+i\infty}{ds\,dt\over(2\pi i)^2}&\Gamma(-t)\Gamma(-t-a-b)\Gamma(s+t+a)\Gamma(s+t+b)\nn\\
&\Gamma\left({\tau\over2}-s\right)\Gamma\left({d-\beta\over2}-s\right)~m^{(J)}(s,t)\,u^s\,v^{-(s+t)}\,,
\eea
where $\beta={\Delta+J\over2}$ and $\tau={\Delta-J\over2}$ are the conformal spin and conformal twist respectively.  

Let us define for future reference  some terms we are going to use along the main sections. We will refer to the subamplitudes $m^{(J)}(s,t)$ expanding \eqref{mellinblockexp} as {\it Mellin blocks}. 
We also define, 
\be
\gamma_{\lambda,a}=\Gamma(\lambda+a)\Gamma(\lambda-a)\,,
\ee 
In applications, we sometimes need the Mellin amplitude with the extra gamma functions removed, and consider the Mellin (inverse) transformation,
\be\label{mellinrepgdress0}
\tilde{g}(u,v)=\int^{\rho-i\infty}_{\rho+i\infty}{ds\,dt\over(2\pi i)^2}~\widetilde{{\cal M}}(s,t)\,u^s\,v^{-(s+t)}\,,
\ee
we will call $\widetilde{\cal M}$ the amputated Mellin amplitude. 
Finally, it is worth noticing that in the representation \eqref{mellinrepg}, crossing $u\leftrightarrow v$ translates simply to $m(s,t,r)=m(r,t,s)$ in Mellin space, where the variables $(s,t,r)$ satisfies the usual kinematical relation for four particles, namely,
\be s+t+r=-\sum_{i=1}^4{\Delta_i\over4}\,, \ee
 where $\Delta_i$ are the conformal weights of the scalar operators involved in the correlator.


\subsection{Primary ``collinear" Mellin blocks}
The leading contribution to the collinear limit $z\to 0$  ($u\to 0$ and $v\to (1-\bz)$) in the {\it coordinate s-channel} \footnote{In this work, whenever we refer to $s-$channel or $t-$channel we mean in cross-ratios space, not in the Mellin variables.} expansion of \eqref{conformalblocks} is controlled by the {\it collinear} conformal blocks \eqref{collinearblocks},
\be\label{blockschannelexp}
u^{{\tau\over2}}k_{\beta}(1-v)=u^{{\tau\over2}} (1-v)^{\beta\over2}{}_2F_1\left({\beta\over 2}+a,{\beta\over 2}+b\,, \beta\,|1-v\right)\,.
\ee
It can be seen that this block can be reproduced from the representation \eqref{mellinrepg} by considering the {\it leading residue} at the pole $s={\tau\over 2}$ from the Mellin amplitude block $m^{(J)}(s,t)$, which in turns can be expressed as \cite{Costa:2012cb, Dolan:2011dv,  Dey:2017fab},
\be\label{MellinAmp2}
m^{(\beta)}(\tau/2,t)=\,{(t+a)_{\tau/2}(t+b)_{\tau/2}\over \kappa_{\beta}}\,P_{\beta/2}(a,b | t)\,,
\ee
with
\be
\kappa_{\beta}={\gamma_{{\beta\over2},a}\gamma_{{\tau\over2},b} \over \Gamma(\beta)\Gamma(\beta-1)}\,,
\ee
and
\be\label{HanhJ} 
P_{\beta/2}(a,b,\tau | s)={(a)_{\beta/2}\,(b)_{\beta/2}\over\Gamma\left({\beta\over2}+1\right)}{}_3F_2\bigg[\begin{matrix} -{\beta\over2},\,~~ -t,\,~~{\beta\over2}-1\\
\ \ a \ \ , \  b
\end{matrix}\,\vert 1\bigg]\,,
\ee
where $(\cdots)_n$ denotes the Pochhammer symbol \footnote{In writing the Mellin block in this form, we have made the assumption that $\beta/2$ is an integer. However, since it is an hypergeometric function, we can assume that there exist an analytic continuation to any other value of $\beta/2$ and we still will use the blocks in the same form. }.
As it was previously noticed by G. Korchemsky \cite{Korchemsky:1994um} and in \cite{Costa:2012cb, Dey:2017fab}, the polynomials $P_{\beta/2}(a,b|t)$ are known in the literature as continuous Hahn polynomials and we have used the definition given at \cite{Andrews}. They satisfy the orthogonality relation \cite{Andrews} 
\bea
&&\int_{-i\infty}^{i\infty}{dt\over 2\pi i}\Gamma(-t)\Gamma(-t-a-b)\Gamma\left(t+{\tau\over2}+a\right)\Gamma\left(t+{\tau\over2}+b\right)P_J(a,b,\tau | t)P_K(a,b,\tau | t)\nn\\
&&={\gamma_{{\beta\over2},a}\gamma_{{\beta\over2},b}\over(\beta-1)\,\Gamma\left(\Delta-1\right)}\delta_{J,K}
\eea
However, the Mellin blocks \eqref{MellinAmp2} are not quite equal to Hanh polynomials due to the $t-$dependent factor in front of $P_{\beta/2}(a,b,\tau | s)$\footnote{It is worth to mention that if we were defined the Mellin blocks in terms of the spin $J$ instead of the conformal spin $\beta$, they would be equal to Hanh polynomials up to a normalization factor.}. Fortunately for us, the Mellin blocks still satisfy a similar orthogonality relation, which we will prove in the appendix and reads,
\bea\label{ortogonality}
&&\int_{-i\infty}^{i\infty}{dt\over 2\pi i}\Gamma(-t)\Gamma(-t-a-b)\Gamma\left(t+{\tau\over2}+a\right)\Gamma\left(t+{\tau\over2}+b\right)\,m^{(\beta)}(\tau/2,t)m^{(\beta')}(\tau/2,t)\nn\\
&&=\,{\Gamma(\beta-1)\over\kappa_{\beta}\,\Gamma\left({\beta\over2}+1\right)}\delta_{\beta,\beta'}
\eea
At the collinear limit, the four point function \eqref{conformalblocksexp} can be expanded as,
\bea\label{conformalblocksexputo0}
{\cal G}(u\to 0,v)&\sim&\sum_{\beta} c_{\tau,\beta}\, u^{{\tau\over2}}k_{\beta}(1-v)\,.
\eea
Therefore we can use the Mellin block  \eqref{MellinAmp2} to write down a corresponding Mellin block decomposition of the Mellin amplitude as,
 \bea\label{mellinatbeta}
{\cal M}(\tau/2,t)&\sim&\sum_{\beta} c_{\tau,\beta}\, m^{(\beta)}( \tau/2,t)\,.
\eea

\subsection{Inverting the Mellin decomposition.}

By using the orthogonality relation \eqref{ortogonality} on \eqref{mellinatbeta}  we get an inversion for the Mellin amplitude expansion  in Mellin space.

\bea\label{mellininversion1}
&&\int_{-i\infty}^{i\infty}{dt\over 2\pi i}\Gamma(-t)\Gamma(-t-a-b)\Gamma\left(t+{\tau\over2}+a\right)\Gamma\left(t+{\tau\over2}+b\right)m^{(\beta)}( \tau/2,t))\,{\cal M}(\tau/2,t)\nn\\
&&=c_{\tau,\beta}\,{\Gamma(\beta-1)\over\kappa_{\beta}\,\Gamma\left({\beta\over2}+1\right)}\,.
\eea

Is worth remarking here that this formula is applicable on the region of validity of the expansion \eqref{conformalblocksexputo0}, namely, for the collinear $s-$channel expansion.  Obviously there is a similar formula valid for the analogous limit on a $t-$channel expansion due to cross symmetry in Mellin space.

\vspace{0.3cm}
{\it Light-cone double-discontinuity}
\vspace{0.2cm}

We are after an equation analogous to  \eqref{simonformula} in Mellin space, and hence we would like to consider an inversion that does not contains contributions from double twist operators. In cross-ratios space, Caron-Huot realized that this is the role played by the light-cone discontinuities of the four point function, which can be isolated by using a clever Froissart-Grivot-like contour deformation. Even thought it would be interesting to see whether or not such a contour deformation exist purely in Mellin space, in this work we will not intend to play a similar game and only are going to use it is as an input.

Sticking at the collinear region, the double-discontinuity across the light-cone branch-cut  $u\to 0$ of the four point function translates to the following in Mellin space,
\bea
{\rm dDisc}({\cal G}(u,v))= \int^{\rho-i\infty}_{\rho+i\infty}{ds\,dt\over(2\pi i)^2}&\Gamma(-t)\Gamma(-t-a-b)\Gamma(s+t+a)\Gamma(s+t+b)\nn\\
&\Gamma\left({\tau\over2}-s\right)\Gamma\left({d-\beta\over2}-s\right)~{\cal M}^{\rm dis}(s,t)\,|u|^s\,v^{-(s+t)}\,,
\eea
where we have absorbed the phase from the discontinuity  into a new object, 
\be
{\cal M}^{\rm dis}(s,t)\equiv{\cal M}(s,t){{\rm sin}^2(\pi\,s)\over \pi^2} \,,
\ee
such that in practice, the actual inversion formula we are going to use is 
\bea\label{mellininversion}
\int_{-i\infty}^{i\infty}{dt\over 2\pi i}&&\Gamma(-t)\Gamma(-t-a-b)\Gamma\left(t+{\tau\over2}+a\right)\Gamma\left(t+{\tau\over2}+b\right)\,m^{(\beta)}(\tau/2,t)\,{\cal M}^{\rm dis}(\tau/2,t)\nn\\
&&~~~=c_{\tau,\beta}\,{\Gamma(\beta-1)\over\kappa_{\beta}\,\Gamma\left({\beta\over2}+1\right)}\,.
\eea
\section{Simplest examples}
In this section we would like to apply the inversion formula discussed in the section above to some simple cases.
\subsection{Vacuum }
It would be convenient to use some examples that allow us to make a first comparison with the expressions at \cite{Caron-Huot:2017vep}. Following \cite{Caron-Huot:2017vep}, lets start with the vacuum contribution. At the collinear limit we want to consider the $t-$channel block,
\be\label{gvac}
f_{\tau'}(v,u)=\left({v\over 1-v}\right)^{{\tau'\over2}+a}(1-v)^{a}\,.
\ee
The prime on $\tau'$ indicates that this corresponds to the twist of the operators expanding the crossed channel. Is easy to see that the corresponding amputated Mellin amplitude reproducing the function \eqref{gvac} is given by,
\be
 \widetilde{\cal M}(\tau'/2+a,t)={\Gamma\left(1-{\tau'\over2}\right)\Gamma\left(t+{\tau'\over2}+a\right)\over \Gamma\left(t+a+1\right)}\,.
\ee
Therefore, the Mellin amplitude associated to the discontinuity is,
\be\label{leadingamputed}
 \widetilde{\cal M}^{\rm dis}(\tau'/2+a,t)={\Gamma\left(1-{\tau'\over2}\right)\Gamma\left(t+{\tau'\over2}+a\right)\over \Gamma\left(t+a+1\right)}{{\rm sin}^2(\pi\,(\tau'/2+a))\over\pi^2}\,.
\ee
Inserting this expression into \eqref{mellininversion} we have,
\bea\label{mellininversionvac}
&&\int_{-i\infty}^{i\infty}{dt\over 2\pi i}\Gamma(-t)\Gamma(-t-a-b)\Gamma\left(t+{\tau'\over2}+a\right)\Gamma\left(t+{\tau'\over2}+b\right)\nn\\
&&\times\,m^{(\beta)}(\tau'/2, t){\Gamma\left(1-{\tau'\over2}\right)\Gamma\left(t+{\tau'\over2}+a\right)\over \Gamma\left(t+a+1\right)}{{\rm sin}^2(\pi\,(\tau'/2+a))\over\pi^2}=\,{\Gamma(\beta-1)\over\kappa_{\beta}\,\Gamma\left({\beta\over2}+1\right)}\,c_{\tau,\beta}\,.
\eea 
After using the series representation for the hypergeometric function ${}_3F_2(...|z)$ in the definition of $m^{(\beta)}(\tau'/2,t)$ and commuting the sum with the contour integral,  it is straightforward to perform the integration to obtain, 

\be\label{ope01}
c_{\tau',\beta}={\Gamma\left({\beta\over2}-a\right)\Gamma\left({\beta\over2}+b\right)\over\Gamma\left(-{\tau'\over2}-a\right)^2\Gamma(\beta-1)}{\Gamma\left({\beta-\tau'\over2}-1\right)\over \Gamma\left({\beta+\tau'\over2}+1\right)}\equiv I^{a,b}_{{\tau'\over2},{\beta\over2}}\,,
\ee
where we have defined $I^{a,b}_{{\tau\over2},{\b\over2}}$ for future reference and to match the notation of \cite{Caron-Huot:2017vep}.

In the s-channel, the function $f(u,v)$  \eqref{gvac} can be expanded in terms of collinear conformal blocks as \cite{Dolan:2000ut}
\be\label{disconectedexp}
f(u,v)=\sum_{k=0}^{\infty}{c_{\Delta,\,k}}\,u^{\Delta}(1-v)^{k}{}_2F_1\left( {\over}\Delta+k,\,\Delta+k,\,
2\Delta+2k\,\vert 1-v\right)\,.
\ee
where the coefficients $c_{\Delta,\,k}$ has been computed in \cite{Heemskerk:2009pn} by solving the bootstrap equation for free fields.  Comparing this last expansion with \eqref{conformalblocksexputo0} and \eqref{mellinatbeta}  it can be also inverted straightforwardly in Mellin space. We can use crossing on \eqref{gvac} to go to the s-channel, but in practice it is equivalent to use \eqref{ope01}  by exchanging ${\tau\over2}\to -\Delta$  (as well as $J\to k$ and $\beta/2\to \Delta+k$). Taking also $a=b=0$ corresponds to equal dimension operators \footnote{I would like to thank Charlotte Sleight for kindly point it me out the distributional character of Mellin amplitudes for disconected correlators and as well as for call to my attention the related work \cite{Bekaert:2016ezc, Taronna:2016ats}}.
After making those replacements, we obtain the result from \cite{Heemskerk:2009pn}
\be \label{vaccoeff}
c^{\rm free}_{\Delta+k}= I^{0,0}_{{-\Delta},{\Delta+k}}={\Gamma(\Delta+k)^2\over\Gamma(\Delta)^2}{\Gamma(2\Delta+k-1)\over \Gamma(k+1)\Gamma(2\Delta+2k+1)}\,.
\ee

\subsection{Double twist operators}
We would like to considering here the OPE coefficients associated to double twist operators with large spin. In order to do so, we are going to use the same strategy followed by \cite{Komargodski:2012ek}.

 As it has been argued in \cite{Komargodski:2012ek}, large spin operators at low twist in the s-channel are controlled by the t-channel block expansion around $u\to 0$ and $v\to 0$. \\
The conformal blocks admit the  expansion \eqref{blockschannelexp} at $u\to 0$, but since the series expansion of the hypergeometric function at \eqref{blockschannelexp} is around $v=1$, in order to explore the region $v\to 0$ we need to make an analytical continuation to $v=0$, which is given by (see for example \cite{ZAMM:ZAMM19660460536}),
\bea\label{hyperident}
{}_2F_1\left(A,B,D\vert 1-v\right)&=&v^{D-A-B}{\Gamma(D)\Gamma(A+B-D)\over\Gamma(B)\Gamma(A)}{}_2F_1\left(D-A,D-B,1-A-B+D\vert v\right)\nn\\
&&+{\Gamma(D)\Gamma(-A-B+D)\over\Gamma(D-B)\Gamma(D-A)}{}_2F_1\left(A,B,1+A+B-D \vert v\right)\,,
\eea 
applying this formula on \eqref{blockschannelexp}, we have at leading order around $(u,v)\to (0,0)$,
\be
g_{\Delta,\tau}=u^{\tau\over2}(1-v)^{\beta\over2}\left({\Gamma(\beta)\Gamma(-a-b)\over\Gamma(\beta/2-b)\Gamma(\beta/2-a)}+{\Gamma(\beta)\Gamma(a+b)\over\Gamma(\beta/2+a)\Gamma(\beta/2+b)}v^{-a-b}\right) 
\ee
one can see that when $a+b=0$, the expansion around $(u,v)\to (0,0)$ develops logarithms  which are associated to the anomalous dimension of double twist operator. Extracting the leading contribution to \eqref{blockschannelexp} at $(u,v)\to (0,0)$ \& $a+b=0$, we obtain,
\be\label{largejlowtblock}
g^{(J)}_{\Delta}(u,v)\sim u^{{\tau\over2}}(1-v)^{\beta\over2}{\Gamma(\beta)\over\Gamma\left({\beta\over 2}+a\right)\Gamma\left({\beta\over 2}-a\right)}(-2\ln v)\,.
\ee
To go to the crossed channel we need to exchange $u\leftrightarrow v$ and multiply by a factor 
\be
{u^{\Delta_0}\over v^{\Delta_0}}\,,
\ee
from here and on the remaining of this subsection we have taken equal dimension scalars for simplicity in the expressions, i.e $a=b=0$.  At leading order in the crossed channel we have,
\be\label{largejlowtblockt}
g^{(J)}_{\Delta}(v,u)\sim u^{\Delta_0}\, v^{{\tau\over2}-{\Delta_0}}{\Gamma(\beta)\over\Gamma\left({\beta\over 2}\right)^2}(-2\ln u)\,.
\ee
As argue by \cite{Komargodski:2012ek}, the leading contribution in the crossed channel expansion of ${\cal G}(v,u)$ is given by the operators with the lowest twist. Ignoring the contribution from the unity we have,
\be\label{largejlowtblockt2}
 {\cal G}(v,u)\sim \, c^{{\rm min}\,\tau}_{\Delta,\tau} \left({v\over 1-v}\right)^{{\tau\over2}-\Delta_0}{\Gamma(\beta)\over\Gamma\left({\beta\over 2}\right)^2}(-2\ln u)
\ee
where we have denote by $c^{{\rm min}\,\tau}_{\Delta,\tau}$ as the OPE coefficient associated to the operator with the lowest twist that can show up on the $t-$channel expansion.

On the direct channel the leading term \eqref{largejlowtblockt2} should admit a collinear expansion of the form \eqref{conformalblocksexputo0}.
By using the fact that we are considering operators with low twist, (hence low anomalous twist), we can expand at first order in the anomalous dimension $u^{{\Delta_0}+\delta\tau/2}\sim{\delta\tau\over2}u^{\Delta_0}\ln u$. All in all at first order, the $s-$channel expansion in the collinear limit can be written as,
\bea\label{conformalblocksexputo01}
{\cal G}(u\to 0,v)&\sim&u^{\Delta_0}\ln u\sum_{\beta} c_{\tau,\beta}\, (1-v)^{\beta\over2}{}_2F_1\left({\beta\over 2}+a,{\beta\over 2}+b,\,\beta\,|1-v\right)\,.
\eea
such us the crossing equation takes the form,
\be\label{largeslowtexp}
c^{{\rm min}\,\tau}_{\Delta,\tau} \left({v\over 1-v}\right)^{{\tau\over2}-\Delta_0}{\Gamma(\beta)\over\Gamma\left({\beta\over 2}\right)^2}=\sum_{\beta} {\delta\tau\over2}c_{\tau,\beta}\, (1-v)^{\beta\over2}{}_2F_1\left({\beta\over 2}+a,{\beta\over 2}+b,\,\beta\,|1-v\right)\,.
\ee

Now it is trivial to invert the above expansion in Mellin space applying \eqref{mellininversion} and by using essentially the same  amputed  Mellin amplitude \eqref{leadingamputed} by making the replacement 
\be
{\tau'\over 2}\to {\tau\over 2}-\Delta_0\,.
\ee
and adding the multiplicative constants,
\be\label{leadingamputed2twist}
 \widetilde{\cal M}^{\rm dis}(\tau'/2+a,t)=c^{{\rm min}\,\tau}_{\Delta,\tau} {\Gamma(\beta)\over\Gamma\left({\beta\over 2}\right)^2}{\Gamma\left(1-{\tau'\over2}\right)\Gamma\left(t+{\tau'\over2}+a\right)\over \Gamma\left(t+a+1\right)}{{\rm sin}^2(\pi\,(\tau'/2+a))\over\pi^2}\,.
\ee
Putting this back into \eqref{mellininversion} and going through the same computation as in the section above, we obtain
\be\label{ope2}
c_{\tau,\beta}{\delta\tau\over2}=c^{{\rm min}\,\tau}_{\Delta,\tau} {\Gamma(\beta)\over\Gamma\left({\beta\over 2}\right)^2} I^{0,0}_{{\tau\over2}-\Delta_0,{\beta\over2}}\,.
\ee
It was additionally argued in \cite{Komargodski:2012ek} that the coefficients $c_{\tau,\beta}$ expanding the direct channel in \eqref{largeslowtexp} should correspond to those ones in the expansion of the vacuum $c^{\rm free}_{\Delta,\,k}$ \eqref{vaccoeff}. 

Putting all those things together we can write, 
\be\label{leadingspinsol}
{\delta\tau\over2}=c^{{\rm min}\,\tau}_{\Delta,\tau} {\Gamma(\beta)\over\Gamma\left({\beta\over 2}\right)^2}\,{I^{0,0}_{{\tau\over2}-\Delta_0,{\beta\over2}}\over I^{0,0}_{{-\Delta},{\Delta+k}}}
\,.
\ee 
By taking the large$-\beta$ on the right hand side of  the above equation, we have the first correction in $1/\beta$ to the twist of large spin double twist operators in terms of the minimal twist operator, as previously done in \cite{Komargodski:2012ek, Caron-Huot:2017vep}. We are going to see this more generally in the next section.
\subsection{Higher order corrections}
It should be possible in principle to compute leading corrections to the result above, as long as it is still possible to have an expansion in terms of collinear blocks. In this section we would like to illustrate how it would works by using a simple example, namely a exchange of a scalar block in the crossed channel. The collinear limit in the crossed channel for a scalar block around $v\to 0$ is approximated by  \cite{Alday:2015ewa},
\bea
g_{J,\Delta}(u\to 0,v)\sim-\ln u {\Gamma(\Delta')\over\Gamma\left({\Delta'\over 2}\right)^2}\,\,{}_2F_1\left({\Delta'\over 2},{\Delta'\over 2};\Delta'-1\,|v\right)\,.
\eea
such as, taking $a=b=0$ for simplicity, the corrections on higher order in  powers of $v$ for the crossing equation looks like,
\bea\label{largesubleading}
&c^{{\rm sub}}_{\Delta',\tau'} \left({v\over 1-v}\right)^{{\Delta'\over2}-\Delta_0}{\Gamma(\Delta')\over\Gamma\left({\Delta'\over 2}\right)^2}\,\,{}_2F_1\left({\Delta'\over 2},{\Delta'\over 2};\Delta'-1\,|v\right)\nn\\
&=\sum_{\beta} {\delta\tau(J)\over2}c^{\rm free}_{\Delta,\,k}\, (1-v)^{\beta\over2}{}_2F_1\left({\beta\over 2},{\beta\over 2}; \beta |1-v\right)\,.
\eea
Where ${\delta\tau(J)}$ is now explicitly understood as a function of $J$. 
A generalization for the amputated Mellin \eqref{leadingamputed},  perhaps naive,  but which reproduce the collinear expansion at the first line of \eqref{largesubleading} is as follows,
\bea
{\cal M}^{\rm sub-dis}({\Delta'}/2,t)&=&c^{{\rm sub}\,\tau}_{\Delta',\tau'}{\Gamma\left(1-{\Delta'\over 2}+\Delta_0\right)\Gamma\left(t+{\Delta'\over2}-\Delta_0\right)\over\Gamma(t+1)\,\pi^2}{\rm sin}^2\left(\pi\left({\Delta'\over2}-\Delta_0\right)\right)\nn\\
&&{\Gamma(\Delta')\over\Gamma\left({\Delta'\over2}\right)^2}{}_3F_1\bigg[\begin{matrix} \Delta'/2,\,~\Delta'/2,\,~~t+\Delta'/2-\Delta_0\\
\ \ \Delta'-1,\,\, {\Delta'\over2}-\Delta_0
\end{matrix}\,\vert 1\bigg]
\eea
We can now invert the expansion on the second line of \eqref{largesubleading} as,
\bea
&&\int_{-i\infty}^{i\infty}{dt\over 2\pi i}\Gamma^2(-t)\Gamma\left(t+{\Delta'\over2}\right)^2\,m^{(\beta)}(\Delta' , t)\,{\cal M}^{\rm sub-dis}(\Delta'/2,t)\nn\\
&&=\,{\delta\tau(J)\over2}c^{\rm free}_{\Delta,\,k}\,{\Gamma(\beta-1)\over\kappa_{\beta}\,\Gamma\left({\beta\over2}+1\right)}\,.
\eea 
By expanding the hypergeometric function in the definition of ${\cal M}^{\rm sub-dis}({\Delta'}/2,t)$ the computation follows essentially the same steps as in the previous sections term by term on the expansion, so each individual terms give us a function $I_{\beta,\tau}$ , more precisely, after performing the contour integration we end up with,
\be
c^{{\rm sub-min}\,\tau}_{\Delta_0,\tau}\e^{i\pi(\Delta-1)}{\Gamma(\Delta')\over\Gamma\left({\Delta'\over2}\right)^2}\sum_{k=0}^{\infty}{\left({\Delta'\over2}\right)^2_k(-1)^k\over k!\, \left({\Delta'}-1\right)_k}\,I^{0,0}_{{\Delta'\over2}-\Delta_0+k,{\beta\over2}}={\delta\tau(J)\over2}c^{\rm free}_{\Delta,\,k}\,.
\ee
This equation can be rewritten in a terms of a hypergeometric function but that form will not be very illuminating either. We can however try a large $\beta$ limit.
By taking $\Delta'\sim 1$, the leading term on a large $\beta$ expansion on the right hand side is,
\bea
{\delta\tau(J)\over2}&=&c^{{\rm sub-min}\,\tau}_{\Delta_0,\tau}{\Gamma(\Delta')\over\Gamma\left({\Delta'\over2}\right)^2}{\Gamma(\Delta_0)^2\over \Gamma\left(\Delta_0-{\Delta'\over2}\right)^2}\sum_{k=0}^{\infty}{\left({\Delta'\over2}\right)^2_k(-1)^k\over k!\, \left({\Delta'}-1\right)_k}\,\left({2\over \beta}\right)^{\Delta'+2k}\nn\\
&=&c^{{\rm sub-min}\,\tau}_{\Delta_0,\tau}{\Gamma(\Delta')\over\Gamma\left({\Delta'\over2}\right)^2}{\Gamma(\Delta_0)^2\over \Gamma\left(\Delta_0-{\Delta'\over2}\right)^2}\left({2\over \beta}\right)^{\Delta'}\,{}_2F_1\left({\Delta'\over2},\,{\Delta'\over2};\,\Delta'-1\,\vert\left({2\over \beta}\right)^{2}\right)\,.\nn\\
\eea
or even better, taking the same limit but this time with ${\Delta_0\over \beta}$ kept fixed lead us to, 
\bea
{\delta\tau(J)\over2}=c^{{\rm sub-min}\,\tau}_{\Delta_0,\tau}{\Gamma(\Delta')\over\Gamma\left({\Delta'\over2}\right)^2}\left({2\over \beta}\right)^{\Delta'}\,{}_2F_1\left({\Delta'\over2},\,{\Delta'\over2};\,\Delta'-1\,\vert\left({2\Delta_0\over \beta}\right)^{2}\right)\,.
\eea
The last result agrees with the re-summation of the leading terms in the large $J$ expansion for the scalar block performed in \cite{Alday:2015ewa}.

\section{Conclusions and outlook}
In this work we have considered the translation of Caron-Huot's OPE inversion formula \eqref{simonformula} from cross-ratios space to Mellin space in the collinear aproximation. In order to do so we have written the Mellin amplitude residue at $s=\tau/2$ in terms of Mellin blocks that in turns can be expresed in terms of orthogonal polynomials approximately equal to Hanh polynomials. We used the aforementioned orthogonality to invert the expansion in Mellin space and introduced by hand the zeros responsible for the cancellation of the poles associated to double twist operators.  Finally we have have checked the inversion in Mellin space on a couple of simple examples withing the region of validity. 

It would be nice to see if an inversion formula can be formulate entirely in Mellin space without any reference to space-time. In particular, it would be interesting to see what is the analogous contour deformation a l\'a Froissart-Gribov in Mellin space, o even better, if there exist a formulation of lorentzian CFT uniquely in terms of Mellin space variables. 

In the derivation of \eqref{simonformula} it is important that the correlator in position space have a good behavior (bounded) at large energies, in order to avoid contributions from infinity after the contour deformation, which probably keeps out the coefficients for operators of lowest spin, $J=0$ and $J=1$. Despite we have not used a contour deformation explicitly, we believe the inversion in Mellin space considered here might be still not valid for the lowest spins, since in essence it is just a translation of \eqref{simonformula}. This and other important features of the inversion \eqref{simonformula} remains to be better understood in Mellin space.

Over the last decade we have gained some important understanding of the Mellin representation for correlation functions in conformal field theory, however, we feel it is still a very unexplored subject that might not only have something else to teach us but can be a potentially useful tool.


\vspace{5mm}
\acknowledgments
\vspace{3mm}
\noindent
I would like to thank to Juan Maldacena, Nima Arkani-Hamed, Simon Caron-Huot, Luis F. Alday, Hugh Osborn, Jake Bourjaily, David McGady and Charlotte Sleight for comments and enlightening discussions. This work is supported in part by the Danish National Research Foundation (DNRF91), ERC Starting Grant (No 757978) and Villum Fonden.

\appendix

\section{Mellin blocks orthogonality}
In the main body of this work we have defined the polynomial Mellin blocks as,
\be
m^{(\beta)}(\tau/2,t)=\,{(t+a)_{\tau/2}(t+b)_{\tau/2}\over \kappa_{\beta}}\,{(a)_{\beta/2}\,(b)_{\beta/2}\over\Gamma\left({\beta\over2}+1\right)}{}_3F_2\bigg[\begin{matrix} -{\beta\over2},\,~~ -t,\,~~{\beta\over2}-1\\
\ \ a \ \ , \  b
\end{matrix}\,\vert 1\bigg]\,,
\ee
with
\be
\kappa_{\beta}={\gamma_{{\beta\over2},a}\gamma_{{\tau\over2},b} \over \Gamma(\beta)\Gamma(\beta-1)}\,,
\ee
In order to prove the orthogonality relation \eqref{ortogonality}  we are going to follow the same route as for the case of Hanh polynomials \cite{0305-4470-18-16-004}, as there, it is sufficient to show that  $m^{(\beta)}(t)$ is orthogonal to one polynomial of each degree less than ${\beta\over2}$ and then check the constant when they are both of the same degree. Lets consider the following  polynomial of degree $\beta'\over2$, which includes the normalization constants,
\be
q^{(\beta')}(\tau/2,t)={(-t-a-b)_{\beta'\over2}\over\kappa_{\beta'}\,\Gamma\left({\beta'\over2}+1\right)}
\ee
\bea
&&\int_{-i\infty}^{i\infty}{dt\over 2\pi i}\Gamma(-t)\Gamma(-t-a-b)\Gamma\left(t+{\tau\over2}+a\right)\Gamma\left(t+{\tau\over2}+b\right)\,m^{(\beta)}(\tau/2,t)\,q^{(\beta')}(\tau/2,t)\nn\\
&&={(a)_{\beta/2}\,(b)_{\beta/2}\over\kappa_{\beta}\Gamma\left({\beta\over2}+1\right)\kappa_{\beta'}\,\Gamma\left({\beta'\over2}+1\right)}\sum_{n=0}^{\beta/2}
{\left(-{\beta\over2}\right)_n\left({\beta\over2}-1\right)_n\over n! (a)_n(b)_n}
\nn\\
&&\times\int_{-i\infty}^{i\infty}{dt\over 2\pi i}\Gamma(-t+n)\Gamma\left(-t-a-b+{\beta'\over2}\right)\Gamma\left(t+a\right)\Gamma\left(t+b\right)\,,
\eea
we can perform the contour integral by means of Barnes first lemma, leading us to,
\be
{\Gamma\left({\beta\over2}+a\right) \Gamma\left({\beta\over2}+b\right) \Gamma\left({\beta'\over2}-a\right) \Gamma\left({\beta'\over2}-b\right) \over\kappa_{\beta} \Gamma({\beta'\over2})\Gamma\left({\beta\over2}+1\right)\kappa_{\beta'}\,\Gamma\left({\beta'\over2}+1\right)}\,{}_2F_1\left(-{\beta\over2},\,{\beta\over2}-1\,\vert {\beta'\over2}\right)\,,
\ee
here we can use Gauss summation formula (sometimes referred as Gauss hypergeometric theorem), namely, 
\be
{}_2F_1\left(a,\,b;\,c\,|1\right)={(c-b)_{-a}\over (c)_{-a}},
\ee
 to end up with,
\be
{\Gamma\left({\beta\over2}+a\right) \Gamma\left({\beta\over2}+b\right) \Gamma\left({\beta'\over2}-a\right) \Gamma\left({\beta\over2}-b\right) \over\kappa_{\beta} \Gamma({\beta'\over2})\Gamma\left({\beta\over2}+1\right)}\,{(-1)^{{\beta\over2}}\left(-{\beta'\over2}\right)_{\beta\over2}\over \left({\beta'\over2}\right)_{\beta\over2}\kappa_{\beta'}\,\Gamma\left({\beta'\over2}+1\right)}\,,
\ee
which vanishes for  $\beta'<\beta$. 
The constant at $\beta'=\beta$ is equal to 
\be 
{\Gamma(\beta-1)\over\kappa_{\beta}\,\Gamma\left({\beta\over2}+1\right)}\,.
\ee
This proves \eqref{ortogonality}.

\bibliographystyle{JHEP}
\bibliography{MIF}
\end{document}